# Finding missing edges and communities in incomplete networks

**Bowen Yan and Steve Gregory**
Department of Computer Science, University of Bristol, Bristol BS8 1UB, England
E-mail: yan@cs.bris.ac.uk +44 117 954 5142

**Abstract**. Many algorithms have been proposed for predicting missing edges in networks, but they do not usually take account of which edges are missing. We focus on networks which have missing edges of the form that is likely to occur in real networks, and compare algorithms that find these missing edges. We also investigate the effect of this kind of missing data on community detection algorithms.

## 1. Introduction

Many complex systems can be described as networks, in which vertices represent individuals, and edges denote relations between pairs of vertices. Researchers analyse these networks in order to understand the features and properties of them, but often neglect to consider the correctness and completeness of the dataset itself. In most cases, they assume that the data are complete and accurate. Most real-world networks represent the results of investigations and experiments, which are often incomplete and inaccurate [1]. For example, in a 1996 survey of sexual behaviour, only 59% of individuals responded to the interview [2].

Missing edges might affect the properties of networks very much. For example, the experimental results of protein-protein interactions have found different properties by using different methods, because 80% of the interactions of proteins are unknown [3]. Also, community detection algorithms might place some vertices in incorrect communities if some edges are missing.

It is not easy to find missing edges in networks because we do not know where they exist. The properties of networks will be affected in different ways by different types of missing edges. Many edge prediction methods have been proposed, but they all consider different issues, and are normally tested only on networks in which edges are missing at random. We believe that it would be beneficial to know how or why edges are missing.

There are several common reasons for an edge to be missing; for example:

1. Edges may be missing at random [4].

2. Edges may be missing because of a limit on the number of neighbours of a vertex: the problem of *right censoring* by vertex degree in social networks [5].

3. Many real-world networks are too large to analyse, so often a sample is extracted from them; for example, by *snowball sampling*, where a crawling algorithm performs (e.g.) a breadth-first search from an initial vertex to collect edges. Since the sample is a subgraph of the whole network, many edges are missing at the periphery.

4. Edges may be missing when a vertex does not have many neighbours: the *cold ends* problem in biological and information networks [6].

In this paper, we define three types of incomplete network based on the first three of the above forms of missing data, and use them for comparing algorithms that find missing edges. We also investigate the effect of the same types of missing data on community detection algorithms.

**2. Related work**

Researchers' attention has increasingly focused on the evolution of networks. Networks are dynamic objects which grow and change, so many new vertices and edges appear in an original network over time. Although the study of missing edges usually focuses on static networks, both problems concentrate on vertex similarity and the structural properties of networks [7].

Many methods have already been used to predict edges in numerous fields; for example, proximity measures that are based on network topological features [8]; supervised learning methods [9]; and relational learning methods [10,11], which consider relational attributes of elements in a relational dataset. These and others are reviewed in the excellent survey by Lü and Zhou [12]. In our paper, we restrict our attention to proximity measures in undirected, unweighted, and unipartite networks.

One of the most popular concepts of proximity between two vertices, structural similarity, is based on the structure information of a network. This concerns the common features that pairs of vertices have: for example, the more common neighbours two vertices have, the greater the chance that they know each other. These methods include *Common neighbours* and *Jaccard coefficient* [8,13,14,15,16,17,18]; or the path information that two vertices share [19,20,21,22,23]. If two vertices do not have any common neighbours, some extended algorithms can be used to calculate their similarity [7,24], taking account of the similarity between the neighbours of two vertices. In Ref. [25], the transferring similarity, which contains all high-order correlations between vertices, has been proposed.

Also, Kossinets [5] has analysed the effect of missing data on social networks via some statistical methods. These network-level statistical properties include mean vertex degree, clustering coefficient, assortativity and average path length. In Ref. [5], Kossinets assumed that the experimental networks were complete first, and removed some vertices and edges randomly for statistical analysis.

The point that network data are not reliable has also been addressed in [26], who emphasized that it is easy to mislead the results of the experiments with incorrect data. Costenbader and Valente [27] have analysed the stability of centrality measures on sampling networks with missing and spurious data. Borgatti et al. [28] have explored the robustness of centrality measures with missing data. Other work includes research on sampling networks with missing data [29].

Understanding the reasons for missing data is a significant goal of edge prediction, because the structure and features of networks depends on the type of missing data. For example, in scale-free networks, the mean degree is unaffected even if there is a large amount of random missing data, but is severely affected by the removal of vertices with high degree [30].

**3. Methodology**

*3.1. Types of incomplete network*

We simulate three types of missing edges in both artificial networks and real-world networks, each with a different cause.

With our first type, edges are missing at the boundary. Generally, this problem is caused by crawling to collect a sample of a network, e.g., by a breadth-first search, so we call it a *crawled network*. With the second type, edges are missing at random, e.g., because of the non-response problem [3,4,5,27], so we name it a *random-deletion network*. With the third type, edges are missing by reason of the fixed-choice problem (right-censoring by vertex degree [5,31]). Since this problem corresponds to limiting the vertex degree, we call it a *limited-degree network*.

*3.2. Testing edge prediction methods*

For our edge prediction experiments, we construct incomplete copies of an initial network as follows:

1. *Crawled network*: We first choose a starting vertex – one of the vertices with the smallest maximum distance from all other vertices – and then use a breadth-first search from the start vertex to collect a sampled network containing the required number of edges. Figure 1(a) shows the famous karate club network [32] while figure 1(b) shows its crawled network with 50% of the original network's edges. This network includes all edges within distance 2 of the starting vertex (shown in red) as well as some of the edges within distance 3, so that the total number of edges is as required.

2. *Random-deletion network*: An edge is chosen randomly and deleted, repeatedly until the required number of edges remain.

3. *Limited-degree network*: At each step, one of the vertices with the maximum degree, *v*, is chosen randomly, and one of its edges {*u*,*v*} is chosen randomly for deletion, irrespective of the degree of *u*.

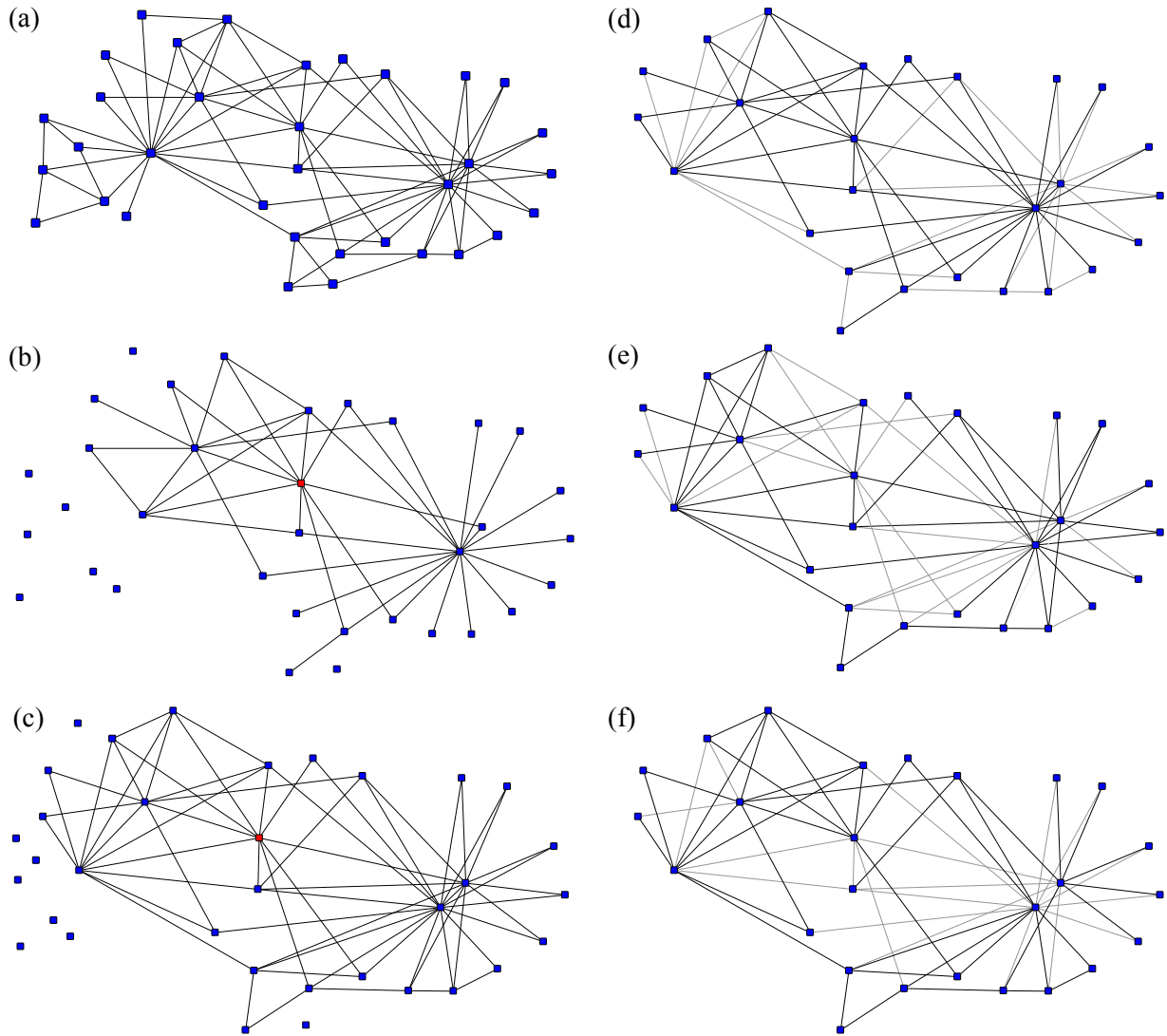

**Figure 1.** (a) Karate network. (b) Crawled karate network, showing missing vertices (those present in the original network but not the crawled network). (c) Induced subnetwork of karate network, showing missing vertices (those present in the original network but not the subnetwork). (d) Crawled karate network, showing missing edges (those present in the induced subnetwork but not the crawled network). (e) Random-deletion karate network, showing missing edges. (f) Limited-degree karate network, showing missing edges.

*3.3. Testing the effect of missing data on community detection algorithms*
The above methods of constructing incomplete networks may cause the network to become disconnected. This is not a problem for our evaluation of edge prediction methods but it would have a severe impact on the comparison of community detection algorithms, because these generally work only on connected networks. Therefore, for our community detection experiments in Section 4.2, we use somewhat different strategies to generate incomplete networks, which ensure that they are all connected and all contain the same set of vertices as well as the same number of edges.

1. *Crawled network*: This is constructed in the same way as above (shown in figure 1(b)). To construct our other incomplete networks, we first construct the subnetwork of the original network induced by the vertices in our crawled network. Figure 1(c) shows the induced subnetwork of the karate network, while figure 1(d) shows the difference between this and the crawled network.

2. *Random-deletion network*: We start with the induced subnetwork (to ensure comparability with the crawled network) and randomly delete edges repeatedly until the required number of edges remain. When choosing an edge to delete, we never choose an edge whose removal would make the network disconnected. Figure 1(e) shows a random-deletion network derived from our induced subnetwork.

3. *Limited-degree network*: Again, we start with the induced subnetwork, and then randomly choose edges {$u,v$} such that $v$ has the maximum degree in this network and the deletion of {$u,v$} would not make the network disconnected. If we cannot find an appropriate $u$, then we choose a $v$ with less than the maximum degree. Figure 1(f) shows a limited-degree network and the induced subnetwork.

*3.4. Network datasets used*

We have experimented with three types of network. First, we use artificial random (Erdős-Rényi) networks [33], because we expect edge prediction methods to be unsuccessful on these.

Second, we use the benchmark networks of Lancichinetti et al. [34]; these are artificial networks that are claimed to reflect the important aspects of real-world networks. The networks have several parameters:

1. $n$ is the number of vertices.
2. $\langle k \rangle$ is the average degree. $k_{max}$ is the maximum degree.
3. $\tau_1$ is the exponent of the power-law distribution of vertex degrees.
4. $\tau_2$ is the exponent of the power-law distribution of community sizes.
5. $\mu$ is the mixing parameter: each vertex shares a fraction $\mu$ of its edges with vertices in other communities.
6. $c_{min}$ is the minimum community size. $c_{max}$ is the maximum community size.

Finally, some real-world networks have been tested, listed in Table 1.

**Table 1.** Real-world networks.

| Name | Ref. | Type | Vertices | Edges |
|---|---|---|---|---|
| scientometrics | [35] | citation | 2678 | 10368 |
| c. elegans | [36] | metabolic | 453 | 2025 |
| email | [37] | social | 1133 | 5451 |
| karate | [32] | social | 34 | 78 |
| terrorists | [38] | social | 62 | 152 |
| grassland species | [39] | food web | 75 | 113 |

## 4. Experiments

*4.1. Edge prediction methods on incomplete networks*

In this section we apply a few edge prediction methods to find the "missing" edges in an incomplete network. We define *score*($u,v$) to be the value of relationship between $u$ and $v$. The higher the score, the more likely they are to be neighbours.

*Common neighbours* (CN). The number of common neighbours that two vertices have is a basic idea that suggests a mutual relationship between them. For example, it may be more likely that two people know each other if they have one or more acquaintances in common in a social network [40]. The function is defined as [8]:

$$score(u,v) = |\Gamma(u) \cap \Gamma(v)|. \tag{1}$$

where $\Gamma(u)$ and $\Gamma(v)$ represent the set of neighbours of vertex $u$ and $v$, respectively.

*Jaccard, Meet/Min, and Geometric*. These three coefficients have a similar definition, related to the probability of triangles in all connected edges of any two vertices [14,16,17,18]. They are defined as [14]:

$$score(u,v) = |\Gamma(u) \cap \Gamma(v)| / |\Gamma(u) \cup \Gamma(v)|. \tag{2}$$

$$score(u,v) = |\Gamma(u) \cap \Gamma(v)| / \min(|\Gamma(u)|, |\Gamma(v)|). \tag{3}$$

$$score(u,v) = |\Gamma(u) \cap \Gamma(v)|^2 / (|\Gamma(u)| \cdot |\Gamma(v)|). \tag{4}$$

*Adamic and Adar* (AA). Adamic and Adar [13] proposed a similarity measure to define the similarity between two vertices in terms of the neighbours of the common neighbours of the two vertices. It is translated into [8]:

$$score(u,v) = \sum_{s \in \Gamma(u) \cap \Gamma(v)} \frac{1}{\log |\Gamma(s)|}. \tag{5}$$

*Resource allocation index* (RA). This is a variant of the method of Adamic and Adar, which assumes that the common neighbours could transmit resources from one vertex to the other one, and is defined as [15]:

$$score(u,v) = \sum_{s \in \Gamma(u) \cap \Gamma(v)} \frac{1}{|\Gamma(s)|}. \tag{6}$$

*Preferential attachment* (PA). This shows that the probability of a connection between two arbitrary vertices is related to the number of neighbours of each vertex. The assumption of this method is that vertices prefer to connect with other vertices with a large number of neighbours. It is described as [40,41]:

$$score(u,v) = |\Gamma(u)| \cdot |\Gamma(v)|. \tag{7}$$

*Hierarchical structure method* (HRG) [42]. This method combines a maximum-likelihood method with a Markov chain Monte Carlo method to sample the hierarchical structure models with probability proportional to their likelihood from the given network. This model is a binary tree with $n$ leaves (the vertices from the given network) and $n$-1 internal nodes, and a probability $p_r$ is associated with each internal node $r$. The probability, $p_r$, of the deepest common ancestor of two vertices represents the probability of a connection between them.

*Stochastic block model method* (BM) [3]. This method uses the stochastic block model, in which vertices are placed into communities and a matrix $Q$ includes the probabilities of connections between communities. $Q_{\alpha\beta}$ is the probability of a connection between two vertices that are in community $\alpha$ and community $\beta$. Then, the probability of an edge can be calculated as explained in Ref. [3].

Because the HRG and BM methods have a higher computational complexity than other methods, we can only evaluate these on some small networks in our experiments reported below.

To determine the accuracy of edge prediction methods, a common measure is the AUC: the area under the ROC (receiver-operating characteristic) curve [42,43,44]. The interpretation of AUC is the probability that the score of a randomly-chosen missing edge is higher than that of a randomly-chosen pair of unconnected vertices.

Figure 2 shows a comparison of edge prediction methods on three incomplete versions of ER networks. Since ER networks are random networks, which do not have any vertex-to-vertex similarity, it is impossible to find missing edges by using edge prediction methods based on common neighbours.

In figure 2, these methods have no effect with any type of missing edges. In contrast, the PA method relies on the number of neighbours of each vertex to find the relation between them. For example, PA performs worse in crawled networks, because the vertices at the boundary of crawled networks always have a low degree; in limited-degree networks, we limit vertices which have a high degree, so PA performs better than other methods on limited-degree networks.

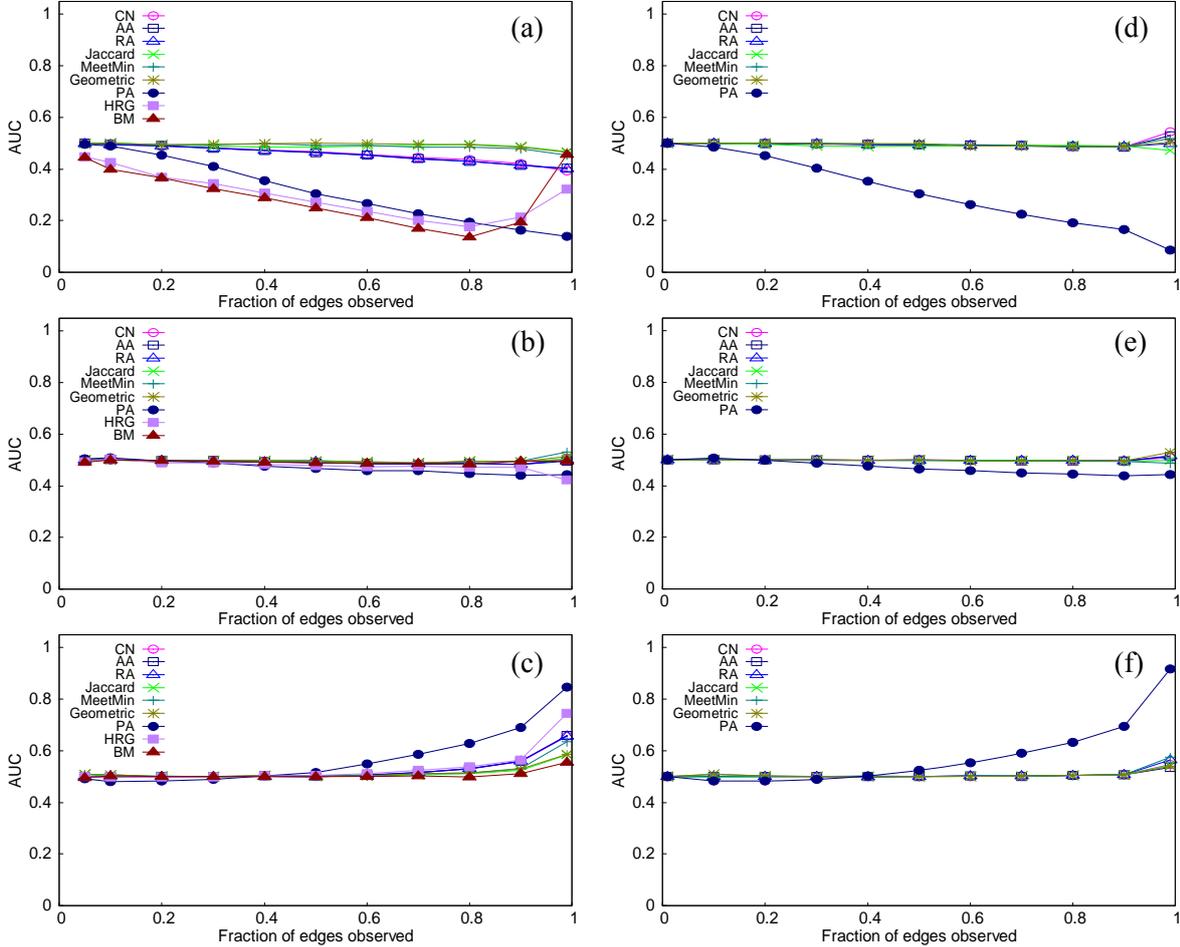

**Figure 2.** Performance of edge prediction methods on incomplete ER networks, with average degree 10. (a) Crawled networks, $n=100$. (b) Random-deletion networks, $n=100$. (c) Limited-degree networks, $n=100$. (d) Crawled networks, $n=1000$. (e) Random-deletion networks, $n=1000$. (f) Limited-degree networks, $n=1000$.

Figures 3-5 show comparisons of edge prediction methods on three incomplete versions of LFR networks (named LFR1, LFR2, and LFR3, respectively), using these parameters:

1. $n = 100$, $\langle k \rangle = 5$, $k_{max} = 15$, $\tau_1 = 2$, $\tau_2 = 1$, $\mu = 0.3$, $c_{min} = 10$, $c_{max} = 20$.
2. $n = 1000$, $\langle k \rangle = 10$, $k_{max} = 25$, $\tau_1 = 2$, $\tau_2 = 1$, $\mu = 0.3$, $c_{min} = 10$, $c_{max} = 20$.
3. $n = 1000$, $\langle k \rangle = 10$, $k_{max} = 25$, $\tau_1 = 2$, $\tau_2 = 1$, $\mu = 0.3$, $c_{min} = 20$, $c_{max} = 40$.

The results of CN, AA, and RA are similar, because they are all based on common neighbours. They perform best on LFR2 networks because these have a higher clustering coefficient than LFR1 and LFR3 (see Table 2). Jaccard, Meet/Min, and Geometric also consider common neighbours, but are affected by other conditions. For example, Jaccard performs best in limited-degree networks because the degrees of many vertices are similar, especially when there are more missing edges. Jaccard is not appropriate when there is a missing edge between a high-degree vertex and a low-degree one.

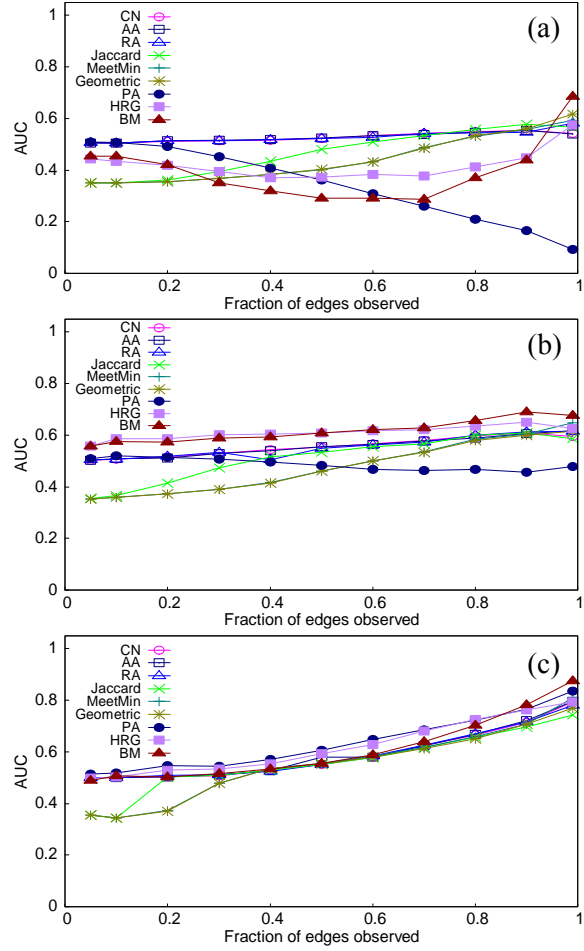

**Figure 3.** Performance of edge prediction methods on incomplete LFR networks: $n$=100, $\langle k \rangle$=5, $k_{max}$=15, $\tau_1$=2, $\tau_2$=1, $\mu$=0.3, $c_{min}$=10, $c_{max}$=20. (a) Crawled networks. (b) Random-deletion networks. (c) Limited-degree networks.

**Table 2.** Clustering coefficient of LFR networks and real-world networks.

| Network | Clustering coefficient |
|---|---|
| LFR1 | 0.1460 |
| LFR2 | 0.2915 |
| LFR3 | 0.1394 |
| karate | 0.2557 |
| terrorists | 0.3609 |
| grassland species | 0.1741 |
| email | 0.1663 |
| scientometrics | 0.0993 |
| c. elegans | 0.1244 |

PA performs well in limited-degree networks, and is not affected by the community size and the clustering coefficient, because PA is appropriate to the mechanism of limited-degree networks. On the contrary, vertices at the periphery usually have a low degree in crawled networks, and it is impossible to know whether the endvertices of missing edges in the crawled and random-deletion networks have high degree.

The HRG and BM methods are better than other edge prediction methods in random-deletion networks, but they do not perform well in crawled networks. In limited-degree networks, these two methods perform similarly to other edge prediction methods.

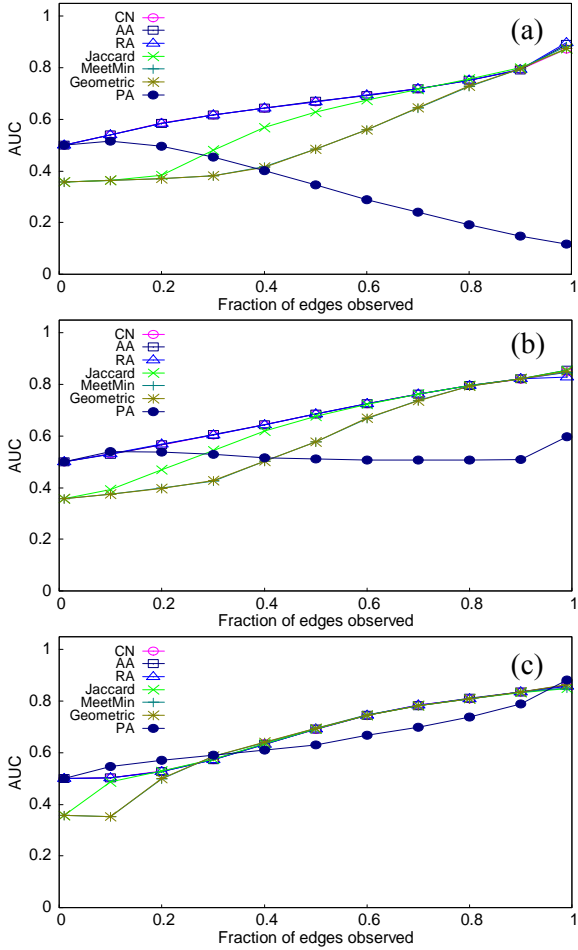 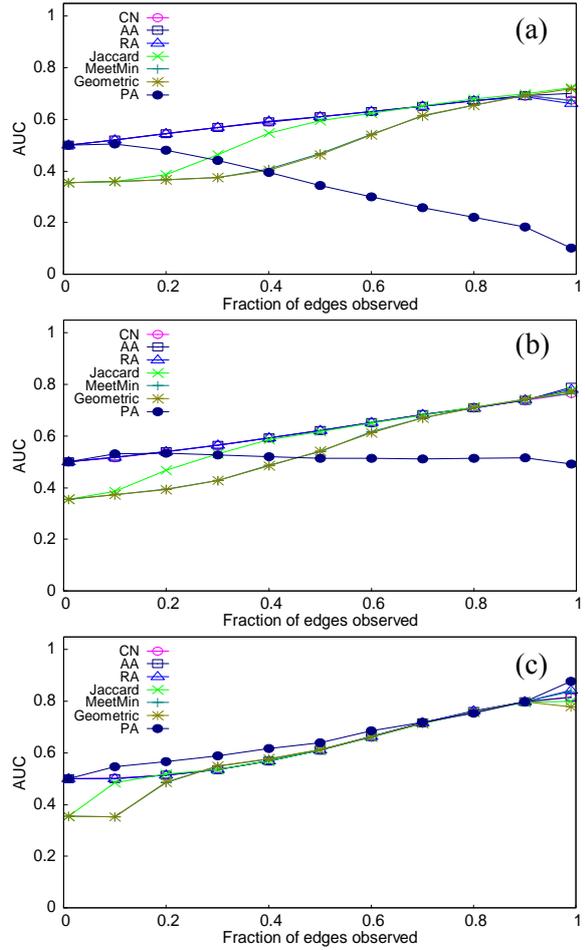

**Figure 4.** Performance of edge prediction methods on incomplete LFR networks: $n=1000$, $\langle k \rangle =10$, $k_{max}=25$, $\tau_1=2$, $\tau_2=1$, $\mu=0.3$, $c_{min}=10$, $c_{max}=20$. (a) Crawled networks. (b) Random-deletion networks. (c) Limited-degree networks.

**Figure 5.** Performance of edge prediction methods on incomplete LFR networks: $n=1000$, $\langle k \rangle =10$, $k_{max}=25$, $\tau_1=2$, $\tau_2=1$, $\mu=0.3$, $c_{min}=20$, $c_{max}=40$. (a) Crawled networks. (b) Random-deletion networks. (c) Limited-degree networks.

Figure 6 shows comparisons of edge prediction methods on three incomplete versions of small real-world networks. Basically, BM and HRG work better than other methods in random-deletion and limited-degree networks. HRG performs well in all types of incomplete networks if the original network is a hierarchical structure (see figures 6(g)-6(i)). PA performs well in limited-degree networks, and sometimes it is even close to or better than BM. CN, RA and AA perform better than BM and HRG in crawled networks if the original network has a high clustering coefficient (see "terrorists" in table 2).

Figure 7 shows comparisons of edge prediction methods on three incomplete versions of larger real-world networks; these are too large for BM and HRG to process in a reasonable time. The results of RA, AA and CN are similar, and they are better than other prediction methods in crawled and random-deletion networks. PA performs well on limited-degree networks.

In general, BM and HRG perform better than other methods in random-deletion and limited-degree networks. However, they are too slow to apply to large networks. In addition to these two methods, CN, RA and AA perform well, while PA usually works well in limited-degree networks. All methods do better on limited-degree networks than on crawled and random-deletion networks.

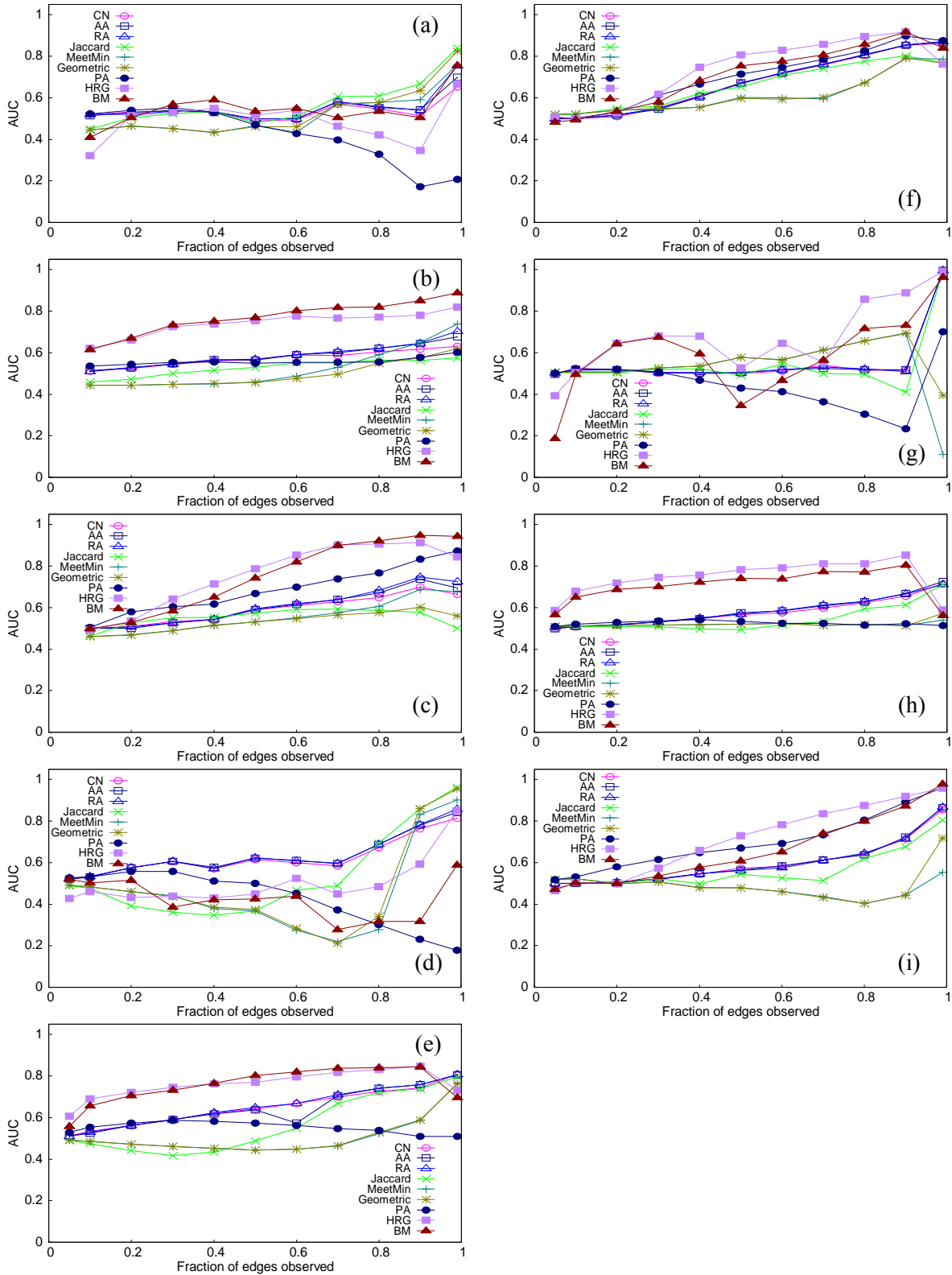

**Figure 6.** Performance of edge prediction methods on incomplete small real networks. (a) Crawled karate network. (b) Random-deletion karate network. (c) Limited-degree karate network. (d) Crawled terrorists network. (e) Random-deletion terrorists network. (f) Limited-degree terrorists network. (g) Crawled grassland species network. (h) Random-deletion grassland species network. (i) Limited-degree grassland species network.

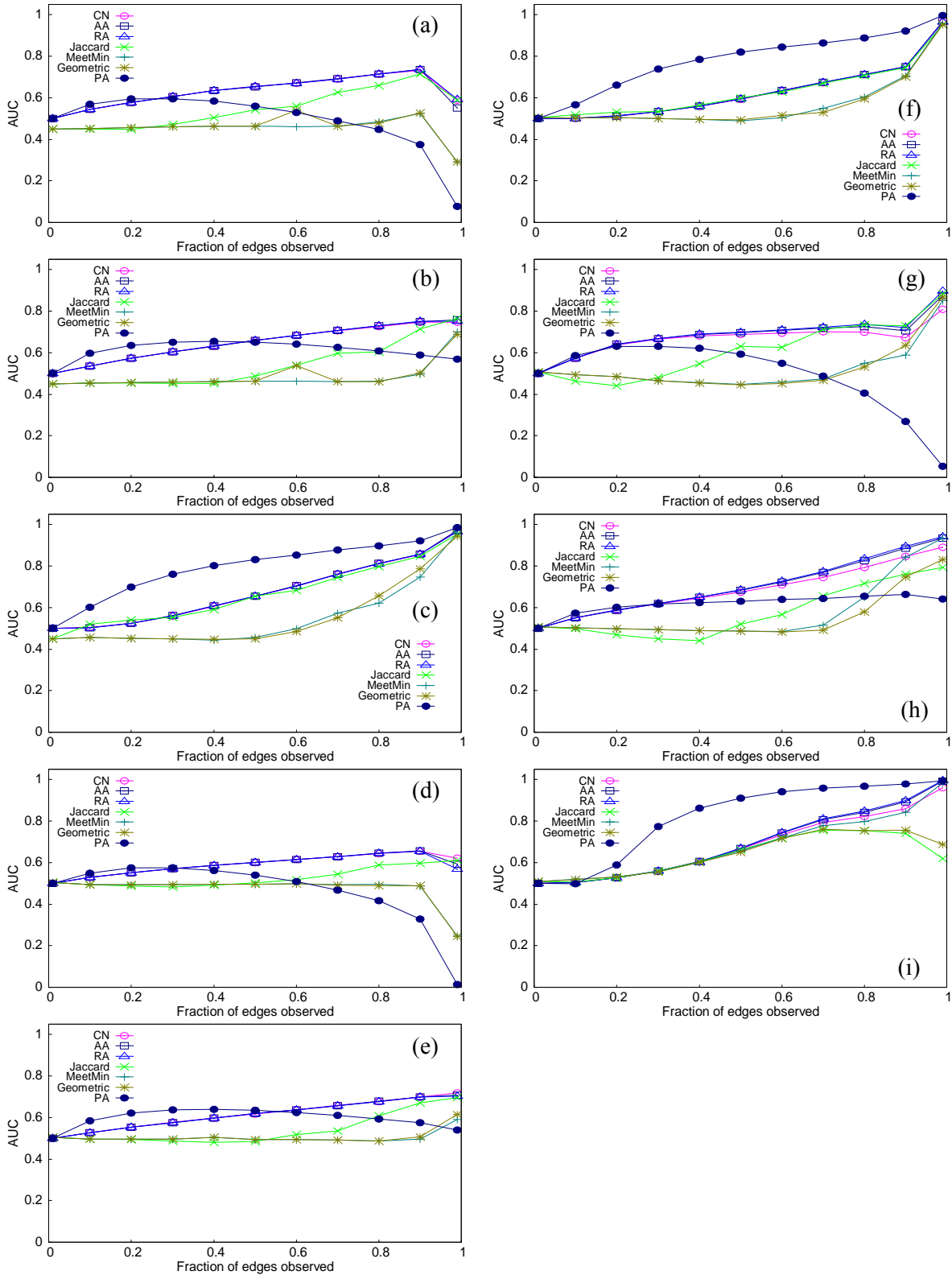

**Figure 7.** Performance of edge prediction methods on incomplete large real networks. (a) Crawled email network. (b) Random-deletion email network. (c) Limited-degree email network. (d) Crawled scientometrics network. (e) Random-deletion scientometrics network. (f) Limited-degree scientometrics network. (g) Crawled c. elegans network. (h) Random-deletion c. elegans network. (i) Limited-degree c. elegans network.

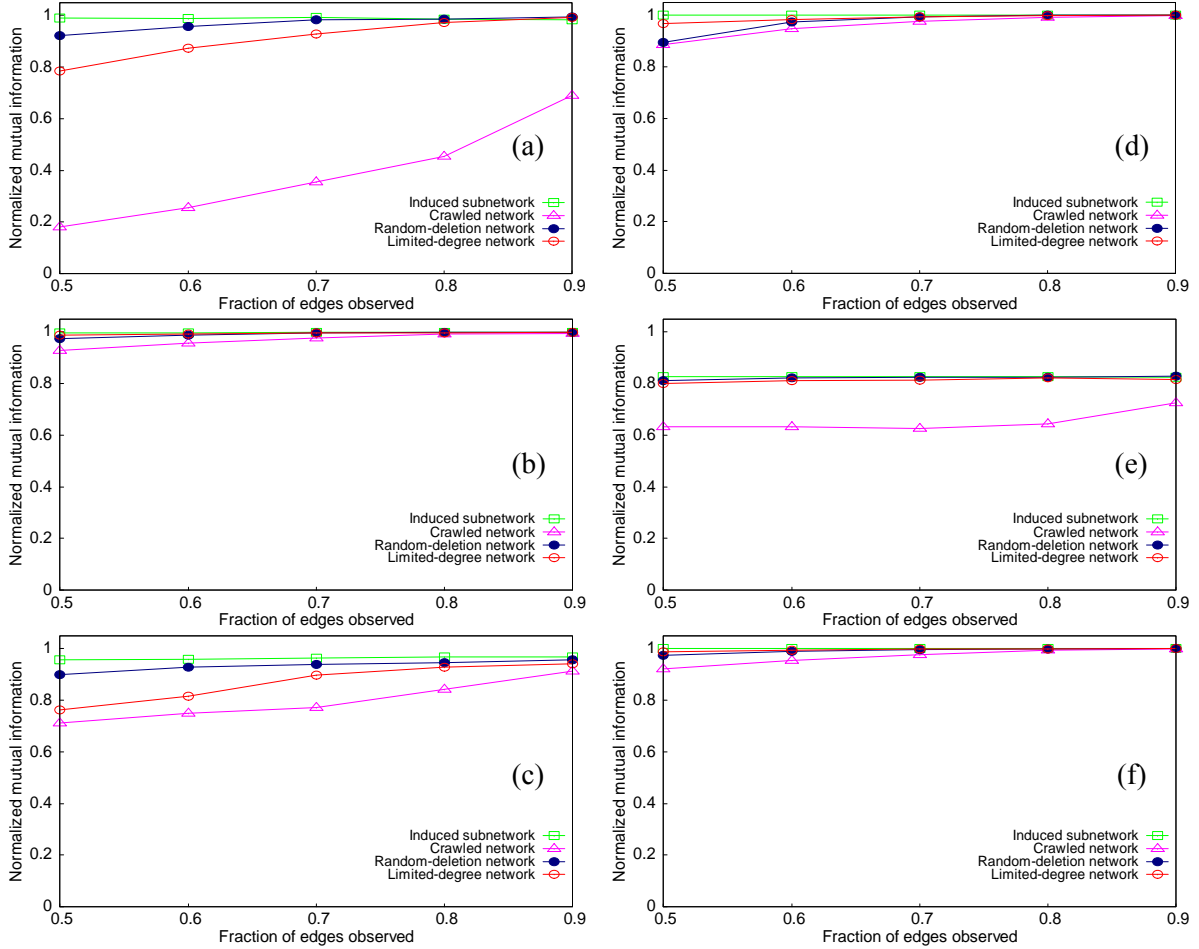

**Figure 8.** Performance of community detection algorithms on various types of incomplete LFR networks. (a) COPRA. (b) CliqueMod. (c) CNM. (d) Walktrap. (e) Wakita and Tsurumi. (f) Louvain method.

*4.2. The effect of missing data on community detection algorithms*
We now investigate the effect of missing data on community detection algorithms: COPRA [45], CliqueMod [46], CNM [47], Wakita and Tsurumi [48], Walktrap [49], and the Louvain method [50]. In this part, we construct the incomplete networks on the induced subnetworks to keep them connected, as explained in Section 3.3, and we compare all of them, including the induced subnetworks, with the original networks. For artificial networks, we use the normalized mutual information (NMI) measure [51] to compare the known partition with the partition found by each algorithm. For real-world networks, since we do not know the real community structure, we use two community detection algorithms (COPRA and the Louvain method), which do not require to be told the number of communities. We run these two algorithms on the original network first, to obtain a partition, and choose the one with the maximum modularity as the "real" partition. Then, we compare the partition obtained by each algorithm for all incomplete networks with that "real" partition.

Figure 8 shows the results of different community detection algorithms on the induced subnetworks and on three incomplete versions of LFR networks (parameters $n=1000$, $\langle k \rangle=20$, $k_{max}=50$, $\tau_1=2$, $\tau_2=1$, $\mu=0.1$, $c_{min}=50$, $c_{max}=100$; the results are averaged over 100 random networks with the same parameters). All of the results of the induced subnetworks are better than those of the incomplete networks. We can also see that the results for crawled networks are usually worse than for other incomplete networks. To understand the reason for this, we calculated the number of removed intercommunity edges, intracommunity edges, and the remaining intercommunity edges in these incomplete networks (see table 3). This shows that random-deletion and limited-degree networks usually lose more intercommunity edges and fewer intracommunity edges than crawled networks. This explains why results are worst for crawled networks. For the other incomplete networks, communities can be found well even when there are many missing edges.

Finally, we examine three real networks: email, scientometrics and c. elegans. Figure 9 shows the performance of COPRA and the Louvain method on induced networks and the three incomplete networks, compared with the original networks. The results are broadly similar to those of artificial networks: the results of random-deleteion and limited-degree networks are better than those of crawled networks.

**Table 3.** Number of intracommunity and intercommunity edges removed in incomplete networks.

| Type of missing edge | % observed edges | # of removed intra edges | # of removed inter edges | # of other inter edges |
|---|---|---|---|---|
| Crawled | 90% | 866 | 85 | 915 |
| Random-deletion | 90% | 852 | 99 | 901 |
| Limited-degree | 90% | 868 | 83 | 917 |
| Crawled | 80% | 1767 | 158 | 842 |
| Random-deletion | 80% | 1726 | 199 | 801 |
| Limited-degree | 80% | 1726 | 199 | 801 |
| Crawled | 70% | 2730 | 298 | 702 |
| Random-deletion | 70% | 2709 | 319 | 681 |
| Limited-degree | 70% | 2716 | 312 | 688 |
| Crawled | 60% | 3400 | 504 | 496 |
| Random-deletion | 60% | 3417 | 487 | 513 |
| Limited-degree | 60% | 3415 | 489 | 511 |
| Crawled | 50% | 4111 | 744 | 256 |
| Random-deletion | 50% | 4067 | 788 | 212 |
| Limited-degree | 50% | 4067 | 788 | 212 |

## 5. Conclusions

Most edge prediction methods have not been evaluated on different types of networks and different forms of missing data. We have found that the performance of these methods is strongly affected by both factors. In particular, we have found that one algorithm, PA, performs well on limited-degree incomplete networks, whereas it has obtained poor results on random-deletion networks in previous research. BM and HRG perform better than other methods on random-deletion and limited-degree networks, even though they are not suitable for large networks. CN, RA, and AA work well in crawled and random-deletion networks. Therefore, to deal with the limited-degree problem in reality, we could choose the PA method, or BM and HRG for small networks. CN, RA, and AA are still good methods for finding other types of missing edges.

We have also investigated the effect of missing data on community detection algorithms. The results show that community detection algorithms perform surprisingly well in the presence of missing edges. The performance depends on the type of missing edges and is worst for crawled networks, probably because of the number of intercommunity edges missing in this type of incomplete network.

## Acknowledgements

We are grateful to Roger Guimerà for sending us the code for the BM method, and we would like to thank the anonymous referees for their useful comments and valuable suggestions.

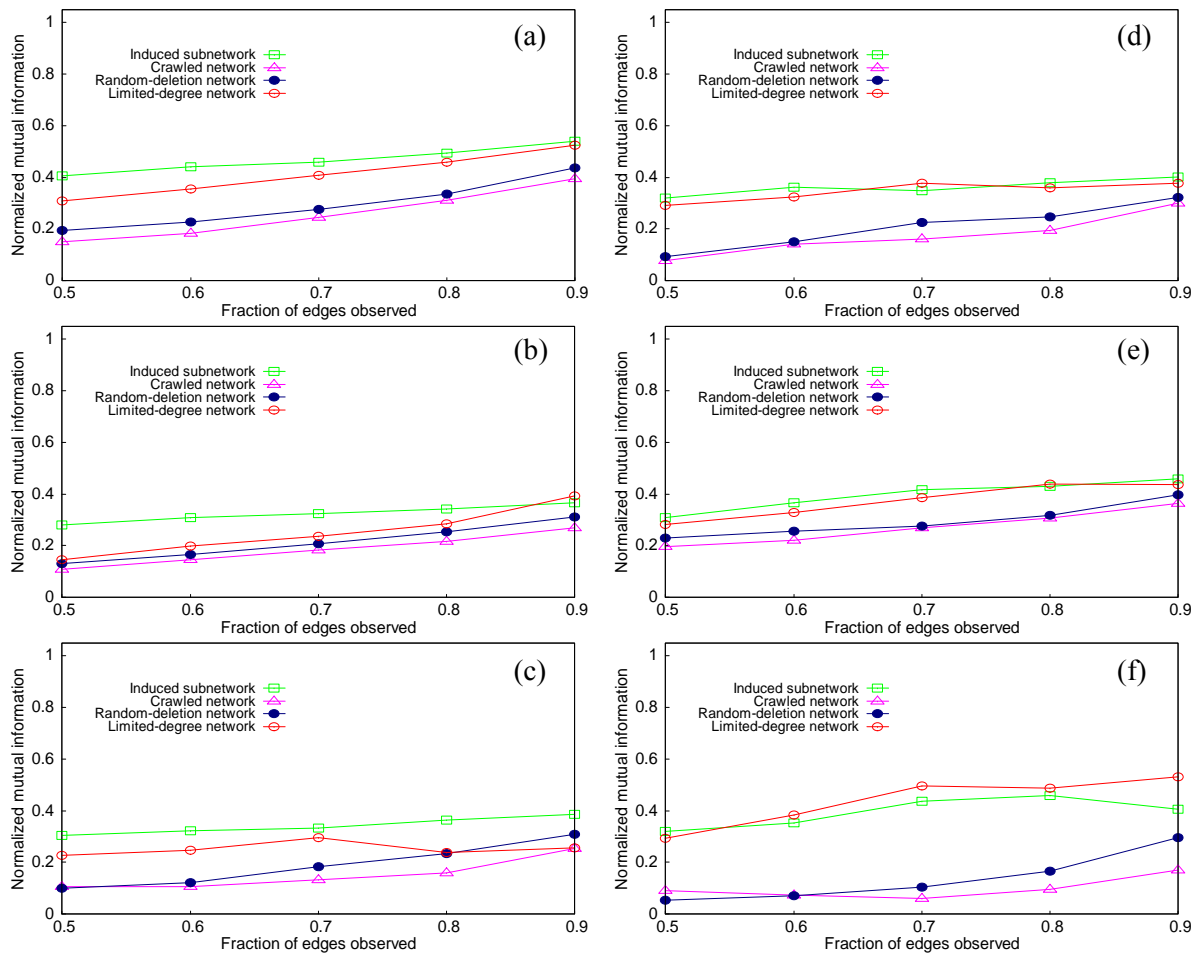

**Figure 9.** Performance of COPRA and the Louvain method on various types of incomplete real networks. (a) Louvain method on incomplete email network. (b) Louvain method on incomplete scientometrics network. (c) Louvain method on incomplete c. elegans network. (d) COPRA on incomplete email network. (e) COPRA on incomplete scientometrics network. (f) COPRA on incomplete c. elegans network.